\documentclass[a4paper]{mem}
\usepackage{natbib}
\usepackage{graphicx}
\begin{document}
   \title{The nature of the Mid-IR faint radio sources from the
     Spitzer First Look Survey
%\thanks{this is a place for a title footnote}
}

   \author{M. Orienti \inst{1,2}, 
   M.A. Garrett \inst{3},\\ 
           C. Reynolds \inst{3}
          \and
          R. Morganti \inst{4}\\
%\thanks{this is a place for placing a footnote in the author field }
}

   \offprints{M. Orienti}
\mail{IRA - CNR, via Gobetti 101, I-40129 Bologna, Italy}

   \institute{Istituto di Radioastronomia - CNR,
via Gobetti 101, I-40129 Bologna, Italy \email{orienti@ira.cnr.it}\\ 
              \and  Dipartimento di Astronomia, Universit\`{a} di
	      Bologna, via Ranzani 1, I-40127 Bologna, 
 Italy\\
              \and Joint Institute for VLBI in Europe, Postbus 2, 7990 AA,
               Dwingeloo, The Netherlands\\
         \and
             Netherlands Foundation for Research in Astronomy, Postbus 2,
7990 AA, Dwingeloo, The Netherlands
             }

   \abstract{
   Data from the Spitzer Space Telescope (the First Look Survey - FLS)
   have recently been made public. We have compared the 24 $\mu$m
   images with very deep WSRT 1.4 GHz observations, centred on the FLS
   verification strip (FLSv). Approximately 75\% of the radio sources
   have corresponding 24 $\mu$m identifications. Such a close
   correspondence is expected, especially at the fainter radio flux
   density levels, where star forming galaxies are thought to dominate
   both the radio and mid-IR source counts. However, a significant
   fracion of radio sources detected by WSRT ($\sim$ 25\%) have no
   mid-IR identification in the FLSv (implying a 24 $\mu$m flux
   density $\leq$ 100 $\mu$Jy). We present initial results on the
   nature of the radio
   sources without Spitzer identification, using data from various
   multi-waveband instruments, including the publicly available R-band
   data from the Kitt Peak 4-m telescope.    
   \keywords{starburst galaxies --
                infrared galaxies --
                low-luminosity AGN
               }
   }
   \authorrunning{M. Orienti et al.}
   \titlerunning{The nature of the Mid-IR faint radio sources in the FLSv}
   \maketitle

\section{Introduction}

Deep radio surveys (S $\leq$ 1 mJy) have clearly indicated the
emergence of a new population of radio sources at mJy and sub-mJy
levels. Several class of objects have been invoked to explain the
steep rise in the integral radio source counts at faint sub-mJy
levels: star forming galaxies, similar to M 82 and Arp 220
(Rowan-Robinson et al. 1993), and low-luminosity AGN like M 84. \\ 
The fact that the locally derived far-IR/radio correlation (e.g. Helou
\& Bicay 1993) also applies to the vast majority of the faint (and
cosmologically distant) radio source population (Garrett 2002),
strongly supports the idea that star forming galaxies begin to
dominate the microJy radio source population.\\
The Spitzer's First Look Survey verification strip, with its 3$\sigma$
sensitivity level of $\sim$ 80 $\mu$Jy at 24 $\mu$m (Marleau et al. 2004),
provides an important opportunity to constrain the nature of the
sub-mJy radio source population.

\section{The samples}
We have extracted a catalogue of sources observed by Spitzer's 
Multiband Imaging
Photometer at 24 $\mu$m (MIPS-24), and using 
the 1.4 GHz WSRT catalogue (1$\sigma$ noise level $\sim$ 8.5
$\mu$Jy; Morganti et al. 2004), we identify two distinct samples:\\

\noindent
- Sample I: 292 radio sources with clear MIPS-24 identification,
  comprising 75\% of the complete FLSv radio sample;\\
- Sample II: 97 radio sources {\it without} MIPS-24
  identification, comprising 25\% of the complete FLSv radio sample.\\

\noindent
Both samples were cross-correlated with the optical R-band FLS
catalogue from the Kitt Peak 4-m telescope (Fadda et al. 2004).

\section{Results} 
Making a comparison between the two radio samples, we find 
different flux density and magnitude distributions: Sample I is
dominated by the faintest radio sources (S $\leq$ 300
$\mu$Jy) with an optical counterpart brighter than R = 22.5, while
Sample II appears to comprise the brighter radio sources (S $>$ 1 mJy)
with optical counterpart fainter than R = 22.5.\\
These results suggest that the two samples are dominated by different
radio source population. This is in agreement with the hypothesis that
the mJy regime is dominated by the faint tail of the AGN
population, while star forming galaxies dominate at sub-mJy levels
(Richards 2000). Our results suggest that the radio sources {\bf
  without} MIPS-24 identification (Sample II) are likely to be
dominated by distant low-luminosity AGN. A study of the SED of variuos
class of objects projected to different redshift supports this
hypothesis: an Ultra-Luminous IR Galaxy like Arp 220 is detectable to
z $\sim$ 0.7 with both WSRT and MIPS-24, while a low-luminosity AGN
like M 84 can be detected up to z $\sim$ 0.7 by WSRT, but only to z
$\sim$ 0.15 by MIPS-24.\\
Another possible explanation is related to the mass and temperature of
the dust in the host galaxy: a galaxy like Arp 220 with a lower
temperature (e.g. $<$ 30 K) is only detectable up to z $\sim$ 0.2 by MIPS-24.\\
Deep spectroscopy, VLBI, sub-mm and X-ray observations will be crucial
in order to further constrain the nature of this class of radio
sources not detected by Spitzer at 24 $\mu$m.\\

\begin{acknowledgements}
This work is based in part on observations made with the {\it Spitzer
  Space Telescope}, which is operated by the JPL, California Institute
  of Technology, under NASA contract 1407. The National Optical
  Astronomy Observatory (NOAO) is operated by the Association of
  Universities for Research in Astronomy (AURA), Inc. under
  cooperative agreement with National Science Foundation.
\end{acknowledgements}

\bibliographystyle{aa}

\end{document}